\documentclass [a4paper,fleqn, 12pt]{article}
\usepackage{graphicx}

\usepackage {amsmath} \usepackage{amssymb} \usepackage{cite}

\newcommand{\pr}{\textbf{Proof.}\ }
\newcommand{\cn}

\begin{document}


\title
{Exact solutions of the Swift-Hohenberg equation with dispersion}

\author
{Nikolay A. Kudryashov, \and Dmitry I. Sinelshchikov}

\date{Department of Applied Mathematics, National Research Nuclear University
MEPHI, 31 Kashirskoe Shosse,
115409 Moscow, Russian Federation}




\maketitle

\begin{abstract}

The Swift-Hohenberg equation with dispersion is considered. Traveling wave solutions of the Swift-Hohenberg equation with dispersion are presented. The classification of these solutions is given. It is shown that the Swift-Hohenberg equation without dispersion has only stationary meromorphic solution.

\end{abstract}






\section{Introduction}

The Swift-Hohenberg equation is one of important equations for description localized structures in the modern physics. This equation occurs in fluid dynamics, optical physics and other fields \cite{Swift,Aranson,Pomeau,Glebsky}. The Swift-Hohenberg equation with dispersion takes the form \cite{Knobloch2006}
\begin{equation}
u_{t}+2\,u_{xx}-\sigma\,u_{xxx}+u_{xxxx}=\alpha\,u+{\beta}\,u^{2}-\gamma\,u^{3},
\label{SH_1}
\end{equation}
where $\sigma$, $\alpha$, $\beta$  and $\gamma$ are parameters of equation. At $\sigma=0$ Eq. \eqref{SH_1} is reduced to the standard Swift-Hohenberg equation.

The wave breaking phenomenon for Eq. \eqref{SH_1} was numerically investigated in \cite{Knobloch2006}. It was shown that localized structures of Eq. \eqref{SH_1} are drift in contrast of structures described by the original Swift-Hohenberg equation.


The aim of this paper is to construct and to classify traveling wave solutions of the Swift-Hohenberg equation with dispersion. We present the traveling wave solutions of the Swift-Hohenberg equation with dispersion and give the classification of all meromorphic traveling wave solutions.  We show that the Swift-Hohenberg equation without dispersion admits only the stationary traveling wave solutions. We also present classification of these stationary solutions as well. To achieve our aim we use the method introduced in  \cite{Kudryashov2010, Kudryashov2010a}.

In recent papers \cite{Kudryashov2010, Kudryashov2010a} two theorems were proved for possible representation of rational, one periodic and doubly periodic solutions in the complex plane of the autonomous nonlinear ordinary differential equations. These theorems allow us to present general forms of one periodic and doubly periodic solutions. At the same time this theorems lead to a new method for constructing meromorphic exact solutions of autonomous nonlinear ordinary differential equations. In the case of one branch of the general solution in the Laurent series we obtain known methods for finding exact solutions, which were developed in the last years \cite{Kudr88, Kudr90, Kudr90a, Kudr92, Parkes01,Fan2000, Fu2002,Parkes02,Kudryashov05,Polyanin,Vernov02,Biswas01,Kudr08a,Kudryashov03,Vitanov,Kudryashov_CNSNS_2010}. In the case of two or more branches for the expansions of the general solution in the Laurent series we obtain new expressions for exact solutions of nonlinear differential equations.

The essence of the approach \cite{Kudryashov2010, Kudryashov2010a} is that the exact solutions of nonlinear ordinary differential equations are found by comparison of the two Laurent series: one of them is the Laurent series  for the general solution of ordinary differential equation and the other is the Laurent series for function which can be a solution to the original equation. Using this approach the meromorphic exact solutions of the third order differential equation were obtained in \cite{Kudryashov2010}. The exact solutions of the Kawahara  and the Bretherton equations expressed via the Weierstrass elliptic function were found in \cite{Kudryashov2010a,KudrPLA11}.

The outline of this paper is the following. In section 2 we study the Swift-Hohenberg equation with dispersion.  We find the traveling wave solutions of this equation and obtain classification of them. In section 3 we consider the exact solution of Eq.\eqref{SH_1} at $\sigma=0$ and show that there is the stationary meromorphic solutions of this equation. We present the classification of these solutions.  In the conclusion we summarize and discuss the results of this paper.

\section{Traveling wave solutions of the Swift-Hohenberg equation with dispersion}

Let us consider the traveling wave solutions of the Swift -- Hohenberg equation with dispersion. Without loss of generality we set $\gamma=-30$ in Eq.\eqref{SH_1}. It is convenient to take this value of the coefficient $\gamma$ for our calculations. In fact, we can obtain this coefficient taking the transformation $u= \left(-\frac{30}{\gamma}\right)^{1/3}\,u^{'}$ into account, where $u^{'}$ is the new variable. In this case Eq.\eqref{SH_1} is reduced to the form
\begin{equation}
u_{t}+2\,u_{xx}+u_{xxxx}-\sigma u_{xxx}=\alpha\,u+\beta\,u^{2}+30\,u^{3}.
\label{SH1}
\end{equation}

Using the traveling wave solutions $u(x,t)=w(z)$, $z=x-C_0\,t$ in \eqref{SH1} we have the nonlinear ordinary differential equation
\begin{equation}
w_{zzzz}-\sigma\,w_{zzz}+2\,w_{zz}-C_{0}\,w_{z}-\alpha\,w-\beta\,w^{2}-30\,w^{3}=0
\label{rSH}
\end{equation}

We suppose that solutions of Eq. \eqref{rSH} has the form of the Laurent series in a neighborhood of the pole $z=z_{0}$
\begin{equation}
w(z)=\sum\limits_{k=0}^{\infty}\,a_{k}\,(z-z_{0})^{k-p}, \quad p>0.
\label{L3}
\end{equation}

Eq.\eqref{rSH} is autonomous and without loss of the generality we can set $z_{0}=0$.

Eq. \eqref{rSH} admits two different formal Laurent expansions in a neighborhood of the second order ($p=2$) pole $z=0$.

Corresponding values of $a_{0}$ and $a_{1}$ are the following
\begin{equation}
a_{0}^{(1,2)}=\pm 2,\quad a_{1}^{(1,2)}=\pm\frac{\sigma}{7}.
\end{equation}

Necessary conditions for existence of elliptic solutions are
\begin{equation}
1) \,\, a_{1}^{(1)}=0,  \qquad 2)\,\,a_{1}^{(2)}=0, \qquad  3)\,\, a_{1}^{(1)}+a_{1}^{(2)}=0.
\label{E3}
\end{equation}

We see that at $\sigma\neq 0$ elliptic solutions can exist only in the third case. At $\sigma=0$ the elliptic solutions can exist in all cases. So at $\sigma\neq 0$ we can look for the elliptic solutions if and only if we take into account both Laurent series for solution of Eq. \eqref{rSH}.

In accordance with classification of meromorphic solutions of autonomous ordinary differential equation given in \cite{Kudryashov2010,Kudryashov2010a}, there are different types of meromorphic exact solutions of Eq. \eqref{rSH}. The first type is the elliptic solutions corresponding to the both Laurent series. The second type is the simply periodic solutions corresponding to one of the Laurent series or to the both Laurent series.

The Fuchs indices corresponding to expansions of the general solution in the Laurent series are the following
\begin{equation}
j_{1}=-1,\quad j_{2}=8, \quad j_{3,4}=\frac{1}{2}\left(7\pm i\sqrt{71}\right).
\end{equation}

We see that the Fuchs indices $j_{2}$ has the positive integer value. So two expansions of the general solution can exist if $a_{8}$ is arbitrary constant.

We have the formal Laurent expansion for the general solution of Eq.\eqref{rSH} corresponding to $a_{0}^{(1)}=2$  in the form
\begin{equation}
\begin{gathered}
w(z)=\frac{2}{z^{2}}+\frac {\sigma}{7z}+\frac{1}{15}-\frac {23{\sigma}^{2}}{2940}-
\frac {\beta}{90}+\vspace{0.2cm}
\\
\\+\left( \frac {3{\sigma}^{3}}{3430}+\frac {C_{0}}{90}-\frac {\sigma}{126} \right) z+\ldots+a_{8}^{(1)}\,z^{6}+\ldots
\label{L_exp_1}
\end{gathered}
\end{equation}

Series \eqref{L_exp_1} corresponds to the general solution of Eq. \eqref{rSH} in the case
\begin{equation}
\begin{gathered}
\frac {\sigma}{26682793200} \left( 3115800\,\sigma^{7}-
50774976\,\sigma^{5}+54096588\,\sigma^{4}C_{0}+\right. \\ \left.+60289110\,{
\sigma}^{3}\alpha-669879\,\beta^{2}\sigma^{3}+222832008\,\sigma^
{3}-401581656\,C_{0}\,\sigma^{2}-\right. \\ \left.-169414560\,\sigma\,\alpha+235298
\,\beta^{3}\sigma+196944426\,\sigma\,{C_{0}}^{2} -203297472\,
\sigma-\right. \\ \left.-31765230\,\sigma\,\alpha\,\beta+1882384\,\beta^{2}\sigma+
338829120\,\alpha\,C_{0}+\right. \\ \left.+338829120\,C_{0}-3764768\,\beta^{2}
C_{0} \right)-\frac{C_{0}^{2}}{225}=0.
\label{C_1}
\end{gathered}
\end{equation}
The last equality is the compatibility condition for existence of the Laurent series \eqref{L_exp_1}. Series \eqref{L_exp_1} does not exist if relation \eqref{C_1} is not satisfied.

The Laurent expansion for the general solution of Eq.\eqref{rSH} at $a_{0}^{(1)}=-2$ can be presented in the form
\begin{equation}
\begin{gathered}
w(z)=-\frac{2}{z^{2}}-\frac {\sigma}{7z}+\frac {23{\sigma}^{2}}{2940}
-\frac{1}{15}-\frac {\beta}{90}+\vspace{0.2cm}\\
\\
+ \left( \frac {\sigma}{126}-{\frac {C_{0}}{90}}\,-\frac {3{\sigma}^{3}}{3430} \right) z
+\ldots+a_{8}^{(2)}z^{6}+\ldots
\label{L_exp_2}
\end{gathered}
\end{equation}

The Laurent series \eqref{L_exp_2} exists and the coefficient $a_{8}^{(2)}$ can be arbitrary constant if there is the compatibility condition for the Laurent expansion \eqref{L_exp_2} in the form
\begin{equation}
\begin{gathered}
-\frac {\sigma}{26682793200} \left( 3115800\,\sigma^{7}-
50774976\,\sigma^{5}+54096588\,\sigma^{4}C_{0}+\right. \\ \left.+60289110\,{
\sigma}^{3}\alpha-669879\,\beta^{2}\sigma^{3}+222832008\,\sigma^
{3}-401581656\,C_{0}\,\sigma^{2}-\right. \\ \left.-169414560\,\sigma\,\alpha-235298
\,\beta^{3}\sigma+196944426\,\sigma\,{C_{0}}^{2} -203297472\,
\sigma+\right. \\ \left.+31765230\,\sigma\,\alpha\,\beta+1882384\,\beta^{2}\sigma+
338829120\,\alpha\,C_{0}+\right. \\ \left.+338829120\,C_{0}-3764768\,\beta^{2}
C_{0} \right)+\frac{C_{0}^{2}}{225}=0.
\label{C_2}
\end{gathered}
\end{equation}

In fact the compatibility conditions \eqref{C_1} and \eqref{C_2} are different.  However in the case of
\begin{equation}
\alpha=\frac{\beta^{2}}{135}
\end{equation}
conditions \eqref{C_1} and \eqref{C_2} are the same.

At $\alpha=\frac{\beta^{2}}{135}$ from \eqref{C_1} and \eqref{C_2} we have
\begin{equation}
\begin{gathered}
C_{0}=\frac {\sigma }{18522(93\,{\sigma}^{2}-56)}\, \Bigg( 54\sigma^{2}(32522-4381\sigma^{2})+5488(\beta^{2}-270)\pm \\
\pm \Bigg[20412\,{\sigma}^{2}
\Big(442123\,{\sigma}^{6} -1593648\,{\sigma}^{2}-1846684\,{\sigma}^{4}-5883136 \Big)+\\+686\,{\beta}^{2} \Big( 11365704\,{\sigma}^{2}+1118799\,{\sigma}^{4}+43904\,{\beta}^{2}-15410304 \Big)+\\+351298031616
 \Bigg]^{1/2} \Bigg)
\label{C0_expression}
\end{gathered}
\end{equation}

Thus Eq. \eqref{rSH} has the elliptic traveling wave solutions if and only if
$\alpha=\frac{\beta^{2}}{135}$ and $C_{0}$ is determined by expression \eqref{C0_expression}.

Let us present the approach for classification of the meromorphic exact solutions for Eq. \eqref{rSH}. For this purpose we use method of constructing general meromorphic solutions for ordinary differential equations \cite{Kudryashov2010,Kudryashov2010a}. The outline of this method is the following.

1) In the first step we construct the formal Laurent series for the general solution of the equation;

2) In the second step we use the general form of the possible elliptic (simple periodic) solution of the equation and find the Laurent expansions for these functions;

3) In the third step we compare the Laurent series for the general solution of the equation, that were found on the first step, with the Laurent expansions of the possible elliptic (simple periodic) solutions, that were found on the second step;

4) In the fourth step we solve the system of algebraic equations obtained on the third step and find the values of parameters of the equation and parameters of the possible elliptic (simple periodic) solutions.

The formal Laurent expansions of the possible elliptic solutions can be found using textbook \cite{Abramowitz} or Maple.

{\textbf{Theorem 1}{ The elliptic solution of Eq. \eqref{rSH} corresponding to both of the Laurent series \eqref{L_exp_1} and \eqref{L_exp_2} exists if there is the following relation between parameters $\beta$ and $\sigma$
\begin{equation}
14\,{(14\,\beta)^2}=  27\, (941\,\sigma^{4}-13384\,\sigma^{2}+43904)
\label{beta_relation}
\end{equation}
 }}
\pr
In accordance with method from Refs. \cite{Kudryashov2010,Kudryashov2010a} we look for the possible elliptic solution of Eq. \eqref{rSH} in the form
\begin{equation}
\begin{gathered}
w(z)=-\frac{1}{2}\left[\frac{\wp'(z,g_{2},g_{3})+B}{\wp(z,g_{2},g_{3})-A}\right]^{2}-
\frac{\sigma}{14}\,\frac{\wp'(z,g_{2},g_{3})+B}{\wp(z,g_{2},g_{3})-A}+\vspace{0.2cm}
+4\,\wp(z,g_{2},g_{3})+h_{0}
\label{Ws}
\end{gathered}
\end{equation}

Comparing expansion \eqref{L_exp_1} with the Laurent expansion of expression \eqref{Ws}, we have
the following values of parameters and invariants
\begin{equation}
\begin{gathered}
h_{0}=\frac {23{\sigma}^{2}-196}{2940}\mp\frac {\sqrt {1843968+39522\,{\sigma}^{4}-562128\,{\sigma
}^{2}}}{5880},\vspace{0.2cm}\\
\\
A=\frac {23\sigma^{2}-196}{5880},\quad B=0, \quad C_{0}=\frac { \left( 38640\,\sigma^{2}-3627\,\sigma^{4}-21952 \right) \sigma}{1372(93\,{\sigma}^{2}-56)}
,\vspace{0.2cm}\\
\\
g_{2}=\frac{2311\sigma^{4}-38024\sigma^{2}+153664}{6914880}, \vspace{0.2cm}\\
\\
g_{3}=-\frac {\left( 23\,{\sigma}^{2}-196 \right)
 \left( 9439\,{\sigma}^{4}-154056\,{\sigma}^{2}+614656 \right)}{203297472000} ,\vspace{0.2cm}\\
\\
\beta= \frac {3\sqrt {1843968+39522\,{\sigma}^{4}-562128\,{\sigma}^
{2}}}{196}
\label{coeffs_1}
\end{gathered}
\end{equation}

Solving the algebraic system of equations on parameters $h_{0},A,B,g_{2},g_{3}, \beta, C_{0}$ we obtain values for constant $a_8$ in the form
\begin{equation}
\begin{gathered}
a_{8}^{(1,2)}=\pm \frac {1147313{\sigma}^{8}}{14344669624320000}\mp\frac {61991{\sigma}^{6}}{
18296772480000}\pm\frac {1621{\sigma}^{
4}}{31116960000}\mp\\
\\
\mp\frac {41{\sigma}^{2}}{119070000}\pm\frac {1}{1215000}.
\label{condtions_1}
\end{gathered}
\end{equation}
Expressions \eqref{condtions_1} are necessary conditions for existence of the Laurent series \eqref{L_exp_1} and \eqref{L_exp_2}. These conditions show us that the elliptic solution of Eq. \eqref{rSH} contains only one arbitrary constant corresponding to Eq. \eqref{rSH}. In this case we can add the arbitrary constant $z_0$ to variable $z$. We have to remember this fact for exact solutions of Eq. \eqref{rSH}.

Taking into account formulae \eqref{Ws} we obtain the elliptic solution of Eq. \eqref{rSH} in the form
\begin{equation}
\begin{gathered}
w(z)=-\frac{1}{2}\left[\frac{\wp'(z,g_{2},g_{3})}{\wp(z,g_{2},g_{3})-\frac {23\sigma^{2}-196}{5880}}\right]^{2}-\frac{\sigma}{14}\,\frac{\wp'(z,g_{1},g_{2})}{\wp(z,g_{2},g_{3})-\frac {23\sigma^{2}-196}{5880}}+\vspace{0.2cm}\\
\\
+4\,\wp(z,g_{2},g_{3})+\frac {23{\sigma}^{2}-196}{2940}-\frac {\sqrt{1843968+39522\,{\sigma}^{4}-562128\,{\sigma}^{2}}}{5880}
\label{SH_gen_el_solution}
\end{gathered}
\end{equation}
where invariants $g_{2},g_{3}$ are defined by relations \eqref{coeffs_1}.
This solution exists if $\beta$ is defined by \eqref{beta_relation}.

This completes the proof. $\Box$

Let us show that elliptic solution \eqref{SH_gen_el_solution} can be reduced to the simple periodic solution at $\sigma=\pm\sqrt{7}$ and $\sigma=\pm \frac{7\sqrt{1513}}{89}$.
The Weierstrass elliptic function is defined by following differential equation
\begin{equation}
\wp'^{2}=4\,\wp^{3}-g_{2}\wp-g_{3}
\label{wp}
\end{equation}
where $g_{2},g_{3}$  and $\beta$ are determined by \eqref{coeffs_1} and \eqref{beta_relation}.

If the right-hand side of Eq. \eqref{wp} has multiple roots then the Weierstrass elliptic function is degenerated to the trigonometric or rational function. It is possible if invariants are defined by \eqref{coeffs_1} and the values of $\sigma$ are the following
\begin{equation}
\sigma=\pm\sqrt{7}, \quad \sigma=\pm \frac{7\sqrt{1513}}{89}
\end{equation}

In the case of $\sigma=\pm\sqrt{7}$ the simple periodic solutions of Eq. \eqref{rSH} take the form
\begin{equation}
\begin{gathered}
w=-\frac{1}{14} \mbox{csch}^{2} \left( \frac{\sqrt{7}z}{14}\right) --\frac{1}{14} \mbox{csch} \left( \frac{\sqrt{7}z}{14} \right)   -\vspace{0.1cm}\\
-\frac{1}{28} \mbox{sech}^{2}  \left( \frac{\sqrt{7}z}{28} \right)-\frac{\sqrt{-154350}}{5880}.
 \label{Peripdic_solution_1}
\end{gathered}
\end{equation}
In this case the parameters $A,B,h_{0},\beta,C_{0}$ can be written as
\begin{equation}
\begin{gathered}
A=-\frac{1}{168}, \quad B=0, \quad h_{0}=-\frac{70+\sqrt{-154350}}{5880},\\
\\
C_{0}=\pm\frac{17\sqrt{7}}{196},\quad \beta=\frac{3\sqrt{-154350}}{196}.
\end{gathered}
\end{equation}

In the case of $\sigma=\pm \frac{7\sqrt{1513}}{89}$ elliptic solution \eqref{SH_gen_el_solution} is reduced to the simple periodic solution in the form
\begin{equation}
\begin{gathered}
w=-\frac {7}{178} \mbox{cosec}^{2}\left( \frac{\sqrt{623}z}{178} \right)-\frac {\sqrt {1513}\sqrt {623}}{15842} \mbox{cosec}\left( \frac{\sqrt{623}z}{178} \right)  +\vspace{0.1cm}\\+\frac{7}{356}\mbox{sec}^{2}\left( \frac{\sqrt{623}z}{356} \right)+\frac{784980-\sqrt{355468050}\sqrt{7921}}{46575480}.
 \label{Periodic_solution_2}
\end{gathered}
\end{equation}
Solution \eqref{Periodic_solution_2} satisfies Eq. \eqref{rSH} when the parameters are the following
\begin{equation}
\begin{gathered}
A=\frac{7}{2136}, \quad B=0, \quad h_{0}=-\frac {\sqrt {355468050}\sqrt {7921}}{46575480}+{\frac {7}{1068}},\vspace{0.1cm}\\
\\
C_{0}=\pm\frac{49\sqrt{1513}}{31684},\quad \beta=\frac {3\sqrt {355468050}\sqrt {7921}}{1552516}.
\end{gathered}
\end{equation}

Consider simple periodic solutions of Eq. \eqref{rSH} corresponding to the Laurent series \eqref{L_exp_1}. The following theorem is valid.

{\textbf{Theorem 2}{The general form of the simple periodic solution of Eq. \eqref{rSH} corresponding to the Laurent series \eqref{L_exp_1} exist if the following relations between parameters $\beta$ and $\sigma$ are hold
\begin{equation}
\begin{gathered}
\beta=\frac{3}{196\sigma}\Bigg[p^{1/3}+\frac{7\sigma^{2} \left( 283\,\sigma^{4}-4592\sigma^{2}+21952 \right)}{p^{1/3}}\Bigg],\vspace{0.2cm}\\
p= \sigma^{3} \left( 41\sigma^{2}-392 \right)  \left( 71\sigma^{2}-392 \right)  \left( 11\sigma^{2}-392 \right)\pm\\ \pm i 30\sqrt{3}\,\sigma^{5}\left(51\sigma^{2}-392\right)\left(31\sigma^{2}-392\right)
\label{beta_relation_1}
\end{gathered}
\end{equation}
or
\begin{equation}
\begin{gathered}
\beta=\frac{3}{98\sigma}\Bigg[q^{1/3}+\frac{49\,\sigma^{2} \left(421\,\sigma^{4}-4664\,\sigma^{2}+12544 \right)}{q^{1/3}}\Bigg],\vspace{0.2cm}\\
q=\pm 30\sigma^{4}\sqrt {3\left(98-15\,\sigma^{2} \right)}  \left( 614656-189336\,{
\sigma}^{2}+14629\,\sigma^{4} \right)  -\\-\sigma^{3} \left( 239\,\sigma^{2}-1568 \right)  \left( 127\,\sigma^{2}-784 \right)  \left( 11\,\sigma^{2}-392 \right)
\label{beta_relation_2}
\end{gathered}
\end{equation}
}}
\pr

The general form of the simple periodic solution can be written in the form \cite{Kudryashov2010,Kudryashov2010a}
\begin{equation}
\begin{gathered}
w(z)=\frac{\pi}{T}\left[\frac{\sigma}{7}\cot\left\{\frac{\pi z}{T}\right\}-2\frac{d}{dz}\cot\left\{\frac{\pi z}{T}\right\}\right]+h_{0}
\label{Simple_periodic}
\end{gathered}
\end{equation}

Also comparing the Laurent expansion for function \eqref{Simple_periodic} with expansion \eqref{L_exp_1} for the general solution of Eq. \eqref{rSH} we obtain
\begin{equation}
\begin{gathered}
T=\pm 28\pi\sqrt{-\frac{1}{\sigma^{2}}}, \quad a_{8}=\frac{\sigma^{8}}{127508174438400}, \quad
h_{0}=\frac{6-\beta}{90}-\frac{41\sigma^{2}}{5880},\vspace{0.1cm}\\
\\
C_{0}=-\frac{\sigma(201\sigma^{2}-1960)}{2744}.
\label{coeffs_3}
\end{gathered}
\end{equation}
Solution \eqref{Simple_periodic} satisfies Eq. \eqref{rSH} if  the parameter $\alpha$ takes the form
\begin{equation}
\begin{gathered}
\alpha={\frac {1}{2963520(343\,\beta-383\,
{\sigma}^{2}-784)}}\,\Bigg(10976\,{\beta}^{2} \left(686\beta-2352-1149\,{\sigma}^{2} \right) +\\+81\,{\sigma}^{2} \left( 9505216-1565928\,{\sigma}^{2}+82933\,{\sigma}^
{4} \right)-697019904\Bigg)
\end{gathered}
\end{equation}
and $\beta$ is defined by \eqref{beta_relation_1}.

Comparing the Laurent series for function \eqref{Simple_periodic} with the Laurent series \eqref{L_exp_1} for the general solution of Eq. \eqref{rSH} we obtain
\begin{equation}
\begin{gathered}
T=\pm\frac{14\pi}{\sqrt{98-15\sigma^{2}}}, \quad
h_{0}=-\frac{\beta+24}{90}-\frac{127\sigma^{2}}{2940}, \\
C_{0}=\frac{\sigma(171\sigma^{2}-980)}{686}, \vspace{0.2cm} \\
 a_{8}=\frac {{\sigma}^{2} \left( 15\,{\sigma}^{2}-196 \right)  \left( 225\,{\sigma}^{4}-2940\,{\sigma}^{2}+19208 \right)}{33205253760}+\frac{1}{5400}
\label{coeffs_3_1}
\end{gathered}
\end{equation}
In this case solution \eqref{Simple_periodic} satisfies Eq. \eqref{rSH} if  the parameter $\alpha$ takes the form
\begin{equation}
\begin{gathered}
\alpha=\frac {1}{185220(7056+343\,\beta-1563\sigma
^{2})}\,\Big(686\beta^{2} \left( 21168+686\beta-4689\sigma^{2} \right) +\\+27\sigma^{2} \left( 101286528-27836424\,\sigma^{2}+2136139\,\sigma^{4} \right)
-1858719744\Big)
\end{gathered}
\end{equation}
and $\beta$ is defined by \eqref{beta_relation_2}.

As result we have the periodic solution of Eq. \eqref{rSH} taking formulae \eqref{Simple_periodic} into account in the following form
\begin{equation}
\begin{gathered}
w(z)=\frac{\pi}{T}\left[\frac{\sigma}{7}\cot\left\{\frac{\pi z}{T}\right\}-2\frac{d}{dz}\cot\left\{\frac{\pi z}{T}\right\}\right]+\frac{6-\beta}{90}-\frac{41\sigma^{2}}{5880},
\label{Simple_periodic1}
\end{gathered}
\end{equation}
where $T$ and $\beta$ are defined by \eqref{coeffs_3} and \eqref{beta_relation_1} or  by \eqref{coeffs_3_1} and \eqref{beta_relation_2} correspondingly.

This completes the proof. $\Box$

The simple periodic solution corresponding to Laurent expansion \eqref{L_exp_2} can be obtained by analogy with solution \eqref{Simple_periodic1}.

\begin{figure}[h] 
  \centering         
    \includegraphics[width=15cm,height=10cm]{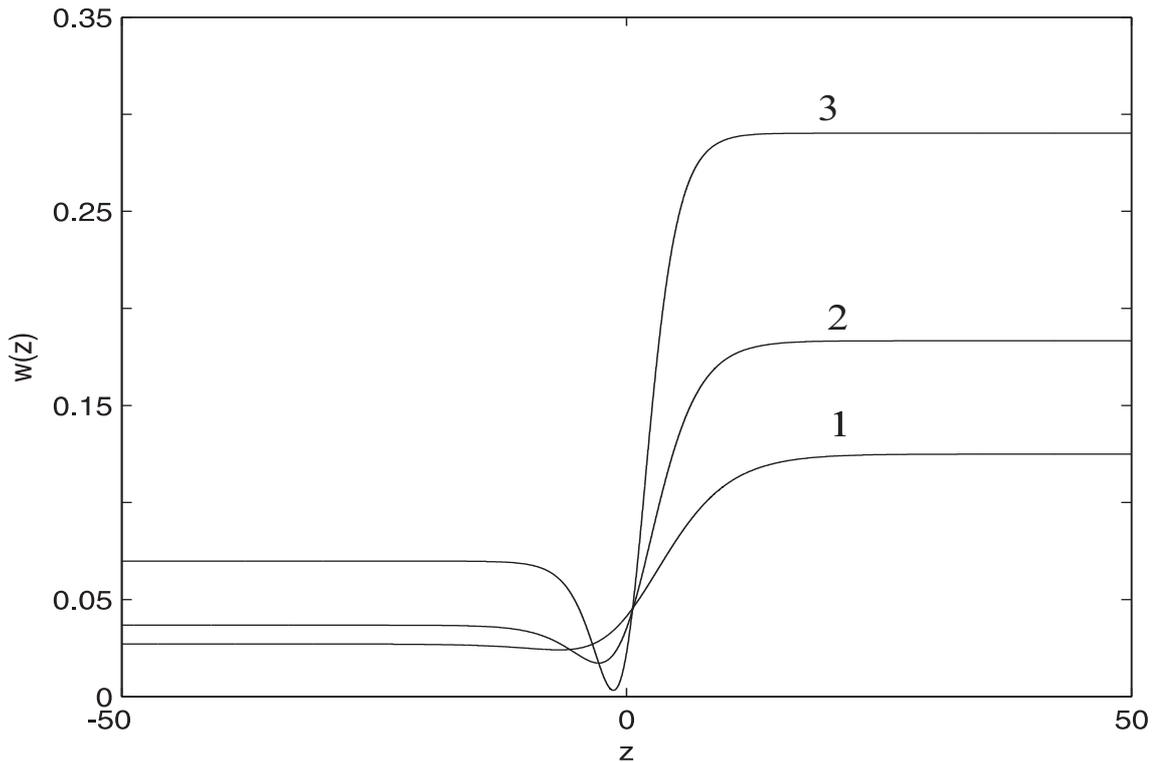}
    \caption{  Exact solution \eqref{Simple_periodic1} of Eq. \eqref{rSH} at $\sigma=2.6,\, 2.65,\, 2.7$ (curves 1, 2, 3).}
  \label{fig:SH_1}
\end{figure}

Let us note that solution \eqref{Simple_periodic1} contains one arbitrary constant corresponding to autonomous of Eq. \eqref{rSH}. If we take this constant purely imaginary then all poles of solution  \eqref{Simple_periodic1} will be on the imaginary axis of the complex plane.  The picture of solution  \eqref{Simple_periodic1} at different values of $\sigma$  on the real axis is demonstrated on  Fig.\ref{fig:SH_1}.

Let us discuss the stability of the simple periodic solution \eqref{Simple_periodic1} of Eq. \eqref{SH1}. To investigate the stability of this solution we use the numerical approach that is the IFRK4 method \cite{Cox,Kassam}. With this aim we have considered the propagation of nonlinear waves described by Eq. \eqref{SH1} with periodic boundary conditions.
To test our numerical approach we use exact solution \eqref{Simple_periodic1}.
This solution is not periodic and we mirror-reflect solution \eqref{Simple_periodic1} with respect to the certain point to obtain the periodic solution in the form of the superposition of exact and reflected solutions as initial conditions. We observe that the relative error of our calculations was less then 5
\% during the time of calculations but the relative error was calculated only on the non-reflected part of initial data. To study the stability of  solution \eqref{Simple_periodic1} we perturb the initial solution using the random noise. The amplitude of the random noise was given between 1\% to 5\% percents of the initial data. We have obtained that the shape of perturbed solutions is not changed at numerical modeling and we believe  that exact solution \eqref{Simple_periodic1} is stable.

\section{Traveling wave solutions of Eq. \eqref{SH_1} at $\sigma=0$.}

Consider the Swift-Hohenberg equation Eq. \eqref{SH_1} at $\sigma=0$.
\begin{equation}
u_{t}+2\,u_{xx}+u_{xxxx}=\alpha\,u+\beta\,u^{2}-\gamma u^{3}
\label{SHH}
\end{equation}
Without loss of generality we take  $\gamma=-30$  in \eqref{SHH} again.  In this case taking into account the traveling wave solutions $u(x,t)=w(z)$, where $z=x-C_0\,t$  in Eq. \eqref{rSH} we obtain  the nonlinear ordinary differential equation
in the form
\begin{equation}
w_{zzzz}+2\,w_{zz}-C_{0}\,w_{z}-\alpha\,w-\beta\,w^{2}-30\,w^{3}=0
\label{rSHH}
\end{equation}

Suppose that there is the expansion of the general solution of Eq. \eqref{rSHH} in the form of the Laurent series  in a neighborhood of the pole $z=z_{0}$
\begin{equation}
w(z)=\sum\limits_{k=0}^{\infty}\,a_{k}\,(z-z_{0})^{k-p}, \quad p>0
\label{LL3}
\end{equation}

As Eq. \eqref{rSHH} is autonomous we can consider that $z_{0}=0$ again.

Substituting \eqref{LL3} into Eq. \eqref{rSHH} we find that $p=2$ and $a_{0}=\pm 2$ again..
The Fuchs indices corresponding of each expansions are the following
\begin{equation}
j_{1}=-1,\quad j_{2}=8, \quad j_{3,4}=\frac{1}{2}\left(7\pm i\sqrt{71}\right)
\end{equation}

Substituting \eqref{LL3} into Eq. \eqref{rSHH} and computing coefficients $a_{k}$  we obtain that the compatibility condition for existence of the Laurent series \eqref{LL3} is the following
\begin{equation}
C_{0}=0.
\end{equation}

As result we obtain that the Swift-Hohenberg equation has only stationary traveling wave solutions.
Let us present classification of the stationary meromorphic exact solutions of Eq. \eqref{rSHH}. For this purpose we use method for constructing general meromorphic solutions of ordinary differential equations from Refs. \cite{Kudryashov2010,Kudryashov2010a} again.

Expansions in the Laurent series of the general solution  of Eq. \eqref{rSHH} in a neighborhood of the  pole $z=0$ are the following
\begin{equation}
w(z)=\pm\frac{2}{z^{2}}-\frac{\beta\mp6}{90}\pm\left(\frac{\beta^{2}-90\alpha-36}{16200}\right)\,z^{2}+\ldots+
a_{8}\,z^{6}+\ldots
\label{SHH_series}
\end{equation}

From expansion \eqref{SHH_series} we see  that the necessary condition for existence of the elliptic solutions is satisfied.

From theorems presented in Refs, \cite{Kudryashov2010,Kudryashov2010a} we obtain that there are some  types of meromorphic solutions of Eq. \eqref{rSHH} : simple periodic solutions (and elliptic solutions)  corresponding to one of Laurent series \eqref{SHH_series} of the general solution  and  simple periodic solutions (and elliptic solutions) corresponding to both of Laurent series \eqref{SHH_series}.

{\textbf{Theorem 3}{The elliptic solutions of Eq. \eqref{rSHH} corresponding to both of the Laurent series \eqref{SHH_series} exist if and only if  $\alpha=\frac{\beta^{2}}{135}$ or  $\beta=0$.}}

\pr
In accordance with method from the references \cite{Kudryashov2010,Kudryashov2010a} we look for the possible elliptic solution of Eq. \eqref{rSHH} in the form
\begin{equation}
w(z)=-\frac{(\wp'(z,g_{2},g_{3})+B)^{2}}{2(\wp(z,g_{2},g_{3})-A)^{2}}+4\wp(z,g_{2},g_{3})+h_{0}.
\label{SH_gen_el}
\end{equation}
Here $\wp(z,g_{2},g_{3})$ is the Weierstrass elliptic function with invariants $g_{2}$, $g_{3}$ and  $\wp'(z,g_{2},g_{3})$ is its derivative with respect to $z$.

Comparing one of expansions \eqref{SHH_series} with the Laurent expansion for expression \eqref{SH_gen_el} we obtain that elliptic solutions exist only if $\alpha=\frac{\beta^{2}}{135}$ or $\beta=0$.

In the case of $\alpha=\frac{\beta^{2}}{135}$ we get the following values of parameters $A,B,h_{0}$
\begin{equation}
\begin{gathered}
A=-\frac{1}{30}, \quad B=0, \quad h_{0}=-\frac{\beta+6}{90}.
\label{s_coeffs_1}
\end{gathered}
\end{equation}
The elliptic solution \eqref{SH_gen_el} of Eq.\eqref{rSHH} takes the form
\begin{equation}
w(z)=-\frac{\wp'^{2}(z,g_{2},g_{3})}{2(\wp(z,g_{2},g_{3})+\frac{1}{30})^{2}}+4\wp(z,g_{2},g_{3})-
\frac{\beta+6}{90},
\label{SHH_gen_el_solution}
\end{equation}
where invariants $g_{2},g_{3}$ are defined by formulae
\begin{equation}
\begin{gathered}
g_{2}=\frac{12-15\alpha}{1620}, \quad g_{3}=\frac{8-25\alpha}{81000}.
\label{Invar_1}
\end{gathered}
\end{equation}
Necessary conditions for series \eqref{SHH_series} to exist is the following
\begin{equation}
\begin{gathered}
a_{8}=\mp \frac{23}{10935000}\mp\frac{11\alpha}{2187000}\mp\frac{7\alpha^{2}}{3499200}.
\label{Invar_2}
\end{gathered}
\end{equation}
At $\beta=0$ parameters $A,B,h_{0}$ can be found from \eqref{s_coeffs_1} at $\beta=0$.
Elliptic solution of Eq. \eqref{rSHH} is expressed by formula \eqref{SHH_gen_el_solution} at $\beta=0$ as well with invariants \eqref{Invar_1} and values $a_8$ from \eqref{Invar_2}. $\Box$

Let us show that elliptic solution \eqref{SHH_gen_el_solution} is reduced to the simple periodic solution in the case of $\beta=\pm\frac{3\sqrt{330}}{5}\,i$.
It is known that the Weierstrass elliptic function is defined by the differential equation
\begin{equation}
\wp'^{2}=4\,\wp^3-g_{2}\,\wp-g_{3}
\label{WP_equation}
\end{equation}
We obtain that the right-hand side of Eq. \eqref{WP_equation} has two multiple root at $\beta=\pm\frac{3\sqrt{330}}{5}\,i$. In this case the Weierstrass elliptic function is reduced to the hyperbolic function.

Solution \eqref{SHH_gen_el_solution} of Eq.\eqref{rSHH} at $\beta=\pm\frac{3\sqrt{330}}{5}\,i$ takes the form
\begin{equation}
w(z)=-\frac{2}{5}\mbox{csch}^{2}\left\{\frac{\sqrt{5}z}{5}\right\}-\frac{1}{5}\mbox{sech}^{2}\left\{\frac{\sqrt{5}z}{10}\right\} \mp\frac{\sqrt{330}}{150}\,i
\end{equation}

The compatibility condition in this case is the following
\begin{equation}
a_{8}=\mp\frac{127}{27000000}
\end{equation}

In the case of $\alpha=\frac{11}{25}$ elliptic solution \eqref{SHH_gen_el_solution} is also reduced to the simple periodic solution of Eq.\eqref{rSHH} in the form
\begin{equation}
w(z)=-\frac{2}{5}\mbox{csch}^{2}\left\{\frac{\sqrt{5}z}{5}\right\}-\frac{1}{5}\mbox{sech}^{2}\left\{\frac{\sqrt{5}z}{10}\right\}
\end{equation}

Let us consider the elliptic solutions of Eq. \eqref{rSHH} which corresponds to one of expansions \eqref{SHH_series}. The following theorem can be formulated.

{\textbf{Theorem 4}{Eq. \eqref{rSHH} has the elliptic solution which corresponds to one of expansions \eqref{SHH_series} at any values of $\alpha$ and $\beta$.}}

\pr

Following the method from the references \cite{Kudryashov2010,Kudryashov2010a} we use the possible elliptic exact solution of Eq. \eqref{rSHH} corresponding to the first Laurent series  \eqref{SHH_series} in the form
\begin{equation}
w=2\,\wp(z,g_{2},g_{3})+h_{0}
\label{SWp}
\end{equation}

Comparing the Laureant series corresponding to function \eqref{SWp} with the first of the expansion \eqref{SHH_series}, we get
\begin{equation}
\begin{gathered}
h_{0}=\frac{6-\beta}{90}, \quad g_{2}=\frac{\beta^{2}-90\alpha-36}{1620},\vspace{0.2cm} \\
\\
g_{3}=-\frac{\beta^{3}+6\beta^{2}-540\alpha-135\alpha\beta-432}{291600}.
\label{coeffs_4}
\end{gathered}
\end{equation}
The compatibility condition for the first Laurent series \eqref{SHH_series} leads to the equality
\begin{equation}
a_{8}=\frac{(36-\beta^{2}+90\alpha)^{2}}{1574640000}
\end{equation}
From \eqref{SWp} as the result of calculations we have
\begin{equation}
w(z)=2\,\wp\left(z,\frac{\beta^{2}-90\alpha-36}{1620},-\frac{\beta^{3}+6\beta^{2}-540\alpha-
135\alpha\beta-432}{291600}\right)+\frac{6-\beta}{90}.
\label{SHH_simple_elliptic_solution}
\end{equation}
This completes the proof. $\Box$

The elliptic solution of Eq. \eqref{rSHH} corresponding to the second expansion from \eqref{SHH_series} can be obtained by analogy with the above presented solution.

Finally let us consider the simple periodic solutions corresponding to one of series \eqref{SHH_series}. The following theorem is hold.

{\textbf{Theorem 5}{Eq. \eqref{rSHH} has the simple periodic solution corresponding to one of expansions \eqref{SHH_series} if
\begin{equation}
\alpha=\frac{(\beta-6)(\beta-24)}{225}
\label{alpha_1}
\end{equation}
or
\begin{equation}
\alpha=\frac { \left(\beta-6 \right)  \left( \beta+12 \right)  \left(
7\beta(\beta+6)\pm(\beta+6)\sqrt{15(\beta+12)(7\beta-12)}+504 \right) }{9 \left(
5(\beta+12)\pm\sqrt{15 \left( \beta+12 \right)  \left( 7\,\beta-
12 \right) } \right) ^{2}}
\label{alpha_2}
\end{equation}
  } }

\pr

The general form of the simple periodic solution corresponding to the first Laurent series  \eqref{SHH_series} can be written \cite{Kudryashov2010,Kudryashov2010a} as
\begin{equation}
w(z)=-\frac{2\pi}{T}\frac{d}{dz}\cot\left\{\frac{\pi z}{T}\right\}+h_{0}.
\label{SP_one_series}
\end{equation}
Comparing the Laurent expansion for function \eqref{SP_one_series} with the first series \eqref{SHH_series} we find the following set of $T, h_{0}$ and $\alpha$ in the form
\begin{equation}
T_{1,2}=\pm\frac{30\pi}{\sqrt{15(6-\beta)}}, \quad h_{0}=0
\label{T1}
\end{equation}
where $\alpha$ is defined by \eqref{alpha_1}. We also have the values
\begin{equation}
\begin{gathered}
T_{3,4}=\pm {\frac {\sqrt {6 \left( \beta+12 \right)  \left( \beta-6
 \right)  \left( 60+5\,\beta+\sqrt {15}\sqrt { \left( \beta+12
 \right)  \left( 7\,\beta-12 \right) } \right) }\pi}{ \left( \beta+12
 \right)  \left( \beta-6 \right) }},\vspace{0.1cm}\\
T_{5,6}=\pm {\frac {\sqrt {-6 \left( \beta+12 \right)  \left( \beta-6
 \right)  \left( 60+5\,\beta+\sqrt {15}\sqrt { \left( \beta+12
 \right)  \left( 7\,\beta-12 \right) } \right) }\pi}{ \left( \beta+12
 \right)  \left( \beta-6 \right) }},\vspace{0.1cm}\\
h_{0}^{(3,4),(5,6)}=-\frac { \left( \beta-6 \right)  \left( \sqrt {15 \left( \beta+12 \right)  \left( 7\,\beta-12\right) }\pm15(\beta+12) \right) }{90(\sqrt {15 \left( \beta+12
 \right)  \left( 7\,\beta-12 \right) }\pm5(\,\beta+12))}
\label{T2}
 \end{gathered}
\end{equation}
where $\alpha$ is defined by \eqref{alpha_2}.

Substituting values of $T$ and $h_{0}$ from \eqref{T1} or \eqref{T2} we obtain the simple periodic solutions of Eq. \eqref{rSHH} at alpha given by \eqref{alpha_1} or \eqref{alpha_2} correspondingly.

This completes the proof. $\Box$

\begin{figure}[h] 
  \centering         
    \includegraphics[width=10cm,height=7cm]{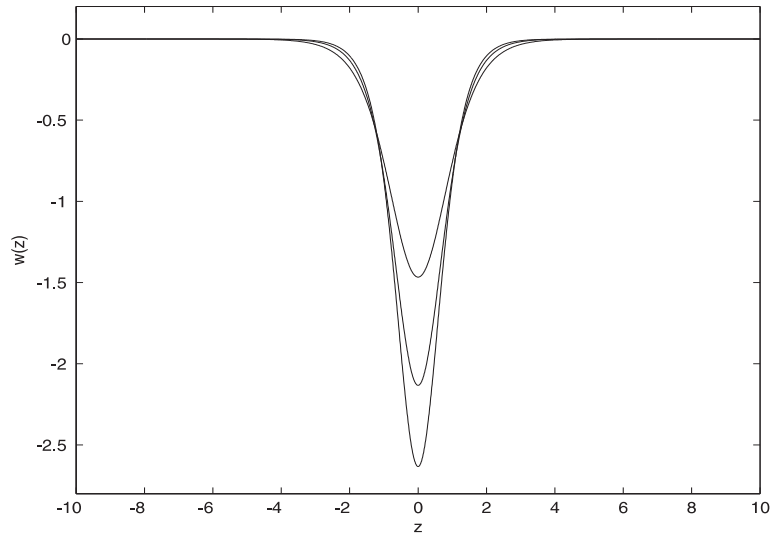}
    \caption{  Exact solution \eqref{SP_one_series} of Eq. \eqref{rSHH} at $\beta=50,\, 70,\, 85$ .}
  \label{fig:SH_2}
\end{figure}

Note that solution \eqref{SP_one_series} contains one arbitrary constant corresponding to autonomous of Eq. \eqref{rSHH}. We can take this constant imaginary then all poles of solution  \eqref{SP_one_series} will be on the imaginary axis of the complex plane. The picture of solution  \eqref{SP_one_series} on the real axis at different values of $\beta$ is demonstrate on the Fig.\ref{fig:SH_2} for this case.

We have investigated the stability of solution \eqref{SP_one_series} by the numerical method again as for solution \eqref{Simple_periodic1}. Solution \eqref{SP_one_series} is periodic and we use \eqref{SP_one_series} as initial data for our numerical calculations. We have perturbed the initial conditions using the random noise with the same amplitude as for solution \eqref{Simple_periodic1}. The results of the numerical simulation showed that the shape of perturbed solutions is not changed at numerical modeling. Thus we believe that exact solution \eqref{SP_one_series} is stable as well.

\section{Conclusion}

In this paper we have studied the traveling wave solutions of the Swift-Hohenberg equation with dispersion. We have shown that nonstationary traveling wave solutions exist in the case of the Swift-Hohenberg equation with dispersion. We have found the elliptic and the simple periodic traveling wave solutions. We have shown that the Swift-Hohenberg equation without dispersion has only the stationary meromorphic traveling wave solutions. We have obtained these exact solutions and presented their classification as well.

\section{Acknowledgements}

This research was  supported by Federal Target Programm
Research and Scientific-Pedagogical Personnel of Innovation
in Russian Federation on 2009-2013.

\end{document}